\documentstyle[preprint,revtex]{aps}
\begin{document}
\draft
\begin{title}
\bf
Is there a renormalization of the 1D conductance \\
in Luttinger Liquid model?
\end{title}
\author{V.V.Ponomarenko}
\begin{instit}
Frontier Research Program, RIKEN, Wako-Shi, Saitama
351-01, JAPAN.  \\
Permanent address: A.F.Ioffe Physical Technical Institute,\\
194021, St. Petersburg, Russia.
\end{instit}

\date{\today}

\begin{abstract}
Properties of 1D transport strongly depend on the proper choice of
boundary conditions. It has been frequently stated that the
Luttinger Liquid (LL)
conductance is renormalized by the interaction as $g \frac{e^2}
{h} $. To contest this result I develop a model of 1D LL wire with the
interaction switching off at the infinities. Its solution shows that
there is no renormalization of the universal conductance while the electrons
have a free behavior in the source and drain reservoirs.

\end{abstract}


\widetext

There is a great interest in the effect of electron-electron interaction
on 1D transport \cite{PhysicsToday}. Following a paper of Apel and Rice
\cite{AR}, Kane and Fisher claimed \cite{KF} that the universal one-channel
conductance $e^2/h$ should be renormalized by a dimensionless constant of the
interaction $g$ in the absence of backscattering. A number of authors
\cite{Others} developed this approach later making it closer to experimental
reality, but keeping this renormalization unchanged. There was serious
doubt, however,  that such a renormalization might occur \cite{DA}. It may be
argued against using a simple Landauer-Buttiker model \cite{BL}, where all
statical  properties of the channel transport should be completely determined
by the numbers and velocities of the electrons flying from the reservoirs
which are connected by the
channel. So, it was not surprising to know that the experimentalists
do not see this renormalization \cite{T}.

In this paper I will describe 1D wire transport with the model accounting
for switching off of the interaction between the electrons inside the
reservoirs. The main result of my solution shows that there is not any
renormalization of the universal conductance whatever the length of the wire
and the way of imposing of the external field .

Such a model could be specified with the Hamiltonian ($\hbar = 1$)
\begin{equation}
{\cal H}= \int dx \{ -iv \psi^+(x)\hat{\sigma_3}
 \partial_x \psi(x) + {1 \over{2}}U
\varphi(x) \rho^2(x) \}
\label{1}
\end{equation}
where the two component field $ \psi_a(x)$ describes the right and left
moving electrons with velocity $v$ through
the only conducting channel between two source and drain
reservoirs. These reservoirs could be modelled as the adiabatic opening of
a number of the channels to the left(right) from $x=0$ ($x=L$), corresponding
to a smooth widening of the potential well there.
Inside the reservoirs electrons become free.
This means that the screened Coulomb
interaction which is a square function of the density
$ \rho(x) = \sum_a \rho_a(x) $
goes to zero for $x<0$ and $x>L$. This switching is ruled by the function
$\varphi(x)$ in Eq. (\ref{1}). Making use of the bosonization technique
\cite{Fr} one could describe such a system with the Langrangian
\begin{equation}
{\cal L}_t={1 \over{8 \pi}} \int dx [ \frac{1}{v}(\partial_t \phi(t,x))^2
-(v+\varphi(x) U/ \pi) (\partial_x \phi(x,t))^2 ]
\label{2}
\end{equation}
The physical quantities could be calculated from the generating functional
\begin{equation}
{\cal Z}= \int D \phi exp\{ iS( \phi )\},
 \ \ \ S( \phi)= \int dt {\cal L}_t(\phi )
\label{3}
\end{equation}
due to connection of $ \phi $ with the electron density $ \rho =
\frac{1}{2 \pi} \partial_x \phi $ and with the electron current
$ j = - \frac{e}{2 \pi} \partial_t \phi $.

To find a current flowing through the channel due to the applied electric
field $ - \partial_y V(t,y) $  I should add
$ {\cal L}_V =\frac{e}{2\pi} \int dy \phi(t,y) \partial_y V(t,y) $
to (\ref{2}) and work out the average of the current with the weight (\ref{3}).
As the action has a Gaussian form the results could be written as
\begin{equation}
<j(t,x)>=-\frac{1}{2\pi}<\partial_t \phi(t,x)>= \int dt' \int dy
\sigma(t-t',x,y) (-\partial_y V(t',y))
\label{4}
\end{equation}
where the conductivity $\sigma(t,x,y)$ is related to the retarded Green
function $ G(t,x,y) $ of the operator
\begin{equation}
\hat{G}^{-1}=- \frac{1}{v^2} \partial_t^2 + \partial_x u^2(x) \partial_x, \ \
u^2(x)=1+U \varphi(x)/(\pi v)
\label{5}
\end{equation}
in the following way: $\sigma(t,x,y)=-\frac{e^2}{\pi v} \partial_t G(t,x,y) $.

First, let me check the result for the case of constant $u$.
Reverse transformation of the Fourier representation
\begin{equation}
\sigma(\omega,k)=\frac{e^2}{\pi u} \frac{i\omega/v'}{(\omega/v')^2-k^2},
\ \ \ \ v'=vu
\label{6}
\end{equation}
gives the expression for the free electron conductivity
\begin{eqnarray}
\sigma(t,x)=\frac{e^2}{2u\pi} \int \frac{d \omega}{2 \pi}\{ \theta(x)
e^{-i \omega (t-x/v')} +\theta(-x) e^{-i \omega (t+x/v')}\} \nonumber \\
=\frac{e^2}{2u\pi} \theta(t) \{ \delta(x-v't)+ \delta(x+v't)\}
\label{7}
\end{eqnarray}
The last line reveals a simple nature of the current in a linear dispersion
approximation
\begin{equation}
<j(t,x)>=-\frac{e^2}{2u\pi} \int dy [\partial_y V(t_{x+}(y),y) +
\partial_y V(t_{x-}(y),y)]
\label{8}
\end{equation}
Since all paths of the particles coming into the point $(t,x)$ from the left
(right) are equivalent $t_{x\pm}(y)=t \mp (x-y)/v' $. The first line of Eq.
(\ref{7}) in agreement with the prescription of Lee and Fisher \cite{LF}
shows that the conductance
$\sigma_0 = \displaystyle \lim_{\omega \rightarrow 0} \sigma(\omega, x) $
at any finite $x$.

Repeating this method in a general case (\ref{5}) one finds
$ \displaystyle \sigma_0 = \frac{e^2}{\pi v}
\lim_{\omega \rightarrow 0} i\omega G(\omega, x,y) $
where the retarded Green function
\begin{equation}
\left[  \frac{\omega^2}{v^2} +\partial_x u^2(x) \partial_x \right]
G(\omega, x,y)=\delta(x-y)
\label{9}
\end{equation}
should be constructed from two proper mutually independent solutions
($ f_{\pm \omega}(x) \propto exp{ \pm i(\omega/v) x}$ at $x \rightarrow \pm
\infty ) $
of the homogenious analog of Eq. (\ref{9}) in a standard
way
\begin{eqnarray}
G(\omega, x,y)=\frac{1}{W(\omega)} [f_{+ \omega}(x)f_{- \omega}(y) \theta(x-y)
+f_{+ \omega}(y)f_{- \omega}(x) \theta(y-x) ] \\
W(\omega)=u^2(x) [ f'_{+ \omega}(x)f_{- \omega}(x)
-f_{+ \omega}(x)f'_{- \omega}(x)]
\label{10}
\end{eqnarray}
Only excitation of the waves of the electron density with small $k$ is
essential. Therefore, I could make a step function approximation to
$u(x)$. Solution of Eq.( \ref{9} ) with this approximation reveals that
$ W(\omega )=2i\omega u( \infty)/v $ at small $ \omega $. Substituting this
into the expression for $ \sigma_0$ I conclude the conductance takes its
universal meaning $ e^2/(2\pi) $ in this model.

Finally, the above considerations prove that the conductance of the 1D wire
is not renormalized by the electron-electron interaction in the absence of
backscattering while electrons have free behavior in the source and drain
reservoirs between which this wire is located.

I am grateful to D. Averin and K. K. Likharev for
elucidative discussions and their
hospitality in Stony Brook, and M. Stopa for his interest and kind help. This
work was supported by the STA Fellowship (Japan) and partially by ONR grant
00014-93-1-0880.

\end{document}